\documentclass[preprint]{iucr}              
\usepackage{xcolor}
 \usepackage{graphicx}
\usepackage{amsmath}
\usepackage{dcolumn}
\usepackage[T1]{fontenc}
\usepackage{breqn}
\usepackage{soul}
\usepackage[utf8]{inputenc} 
     \journalcode{J}              

\usepackage[noabbrev,capitalise]{cleveref}
\usepackage{soul}

\begin{document}                  
\title{On the analysis of two-time correlation functions: equilibrium vs non-equilibrium systems}

\author[a]{Anastasia}{Ragulskaya}{}{}
\author[a]{Vladimir}{Starostin}{}{}
\cauthor[a]{Fajun}{Zhang}{fajun.zhang@uni-tuebingen.de}{}
\author[b]{Christian}{Gutt}{}{}
\cauthor[a]{Frank}{Schreiber}{frank.schreiber@uni-tuebingen.de}{}

\aff[a]{~Institute of Applied Physics, University of Tübingen, Auf der Morgenstelle 10, 72076, \city{Tübingen}, \country{Germany}}
\aff[b]{~Department of Physics, University of Siegen, Emmy-Noether-Campus, Walter-Flex-Str. 3, 57076, \city{Siegen}, \country{Germany}}

\shortauthor{Ragulskaya, Anastasia et. al.}


\maketitle                        

\begin{synopsis}
This article explores widely used approaches for TTC calculations and common methods for extracting relevant information from correlation functions. The results are applicable to a wide range of processes including growth and coarsening.
\end{synopsis}

\begin{abstract}

X-ray photon correlation spectroscopy  (XPCS) is a powerful tool for the investigation of dynamics covering a broad range of time and length scales. The two-time correlation function (TTC) is commonly used to track non-equilibrium dynamical evolution in XPCS measurements, followed by the extraction of one-time correlations. While the theoretical foundation for the quantitative analysis of TTCs is primarily established for equilibrium systems, where key parameters such as diffusion remain constant, non-equilibrium systems pose a unique challenge. In such systems, different projections ("cuts") of the TTC may lead to divergent results if the underlying fundamental parameters themselves are subject to temporal variations. This article explores widely used approaches for TTC calculations and common methods for extracting relevant information from correlation functions on case studies, particularly in the light of comparing dynamics in equilibrium and non-equilibrium systems.  

\end{abstract}

\section{Introduction}
X-ray Photon Correlation Spectroscopy (XPCS) is a highly versatile experimental technique that is widely used to study the dynamics of both soft and hard condensed matter. The current state of XPCS and light sources enables the investigation of the dynamics across an unprecedented range of time scales, from femtoseconds to hours, and length scales that span from microns down to angstroms \cite{Shpyrko2014, Lehmkuhler2021}. XPCS is employed in various areas of condensed matter research to explore the dynamics of colloids \cite{Westermeier2012, Kwasniewski_2014, Angelini2014, ANGELINI2015, Liu2021}, liquids and liquid crystals \cite{Seydel2001, Lu2008, Madsen_2003,vantZand2012}, polymers \cite{Narayanan2007, Conrad2015, NOGALES2016}, metallic and molecular glasses \cite{Leitner2012, Ruta2012, Ruta2013}, proteins \cite{Begam2021, Girelli2021, Ragulskaya_2021_JPhysChemLett, Reiser2022-pj, Chushkin2022}, magnetic systems \cite{Shpyrko2014, Zhang2017}, and clays \cite{Bandyopadhyay_2004}.

The dynamics of a system under investigation is revealed by analyzing the temporal correlations of the scattered intensity. Under coherent illumination, the resulting far-field pattern of the scattered intensity exhibits spots of constructive and destructive interference known as speckles. The dynamics of the sample leads to changes in the speckle pattern. The XPCS technique exploits the fluctuations of these speckles to extract information about the dynamic behavior of the sample. A comprehensive summary of XPCS research studies and future prospects can be found in several reviews such as Shpyrko \cite{Shpyrko2014}, Grübel \cite{Grubel}, Sutton \cite{SUTTON2008}, Lehmkühler \cite{Lehmkuhler2021}, Gutt\& Perakis \cite{Perakis_2020}, and Sinha \cite{Sinha_2014_XPCS_surface}. An overview on the qualitative analysis was presented by Bikondoa \cite{Bikondoa2017}. Nevertheless, a satisfying link between the theory derived for equilibrium systems (including the estimation of physical parameters of the system such as diffusion, viscosity, etc.) and the analogous quantitative analysis of non-equilibrium systems, where the underlying physical parameters evolve with time, is still missing and indeed difficult to achieve. 

The subject of non-equilibrium is, of course, not limited to a specific technique, but increasingly relevant in various fields, for example, glass physics and mode coupling theory \cite{Gotze1999, Martinez2010}, as well as growth phenomena \cite{Headrick2019, Stephenson2019, Dax2023}. Time-resolved correlations have also been pioneered in dynamic light scattering with important insights into temporal heterogeneities, higher-order correlations, and spatial-temporal correlations \cite{Cipelletti_2003, Duri_2006,Duri_2005, Cipelletti_1999} and references therein. The present paper attempts to provide first an overview of the conventional data analysis of the XPCS studies and then to complement previous studies by discussing the quantitative analysis in the light of the connection between equilibrium and non-equilibrium including the comparison based on specific case studies. Although the findings are limited to specific conditions and may not be readily generalized, we hope to inspire further theoretical and numerical investigations to shed light on this issue in a broader context, which is currently underrepresented in the literature. 

\section{Two-time correlation function (TTC).}

\subsection{Conventional calculations of TTC and their connection.}

Data analysis is a crucial step in XPCS, as it involves extracting the relevant information from the correlation functions of the measured time-resolved 2D speckle patterns ($I(t)$). To follow non-equilibrium dynamical evolution during the XPCS measurement, it is customary to use the two-time correlation function (TTC) \cite{Madsen_2010, Bikondoa2017, Sutton2003}:

\begin{equation}\label{eq:TTC_mean}
Corr(\vec{q}, t_1, t_2)=\frac{\langle I(\vec{q}, t_1) I(\vec{q}, t_2) \rangle}{\langle I(\vec{q}, t_1) \rangle \langle I(\vec{q}, t_2) \rangle}.
\end{equation}
This "$Corr$-TTC" calculates the correlation between intensities at times $t_1$ and $t_2$ averaged over all pixels at the same $q$-ring. For simplicity, we shall assume an isotropic sample. Here, the normalization is performed by the mean intensity.

Another possibility for the calculation of the TTC is $G$-TTC - the autocovariance of the intensity normalized by its standard deviation $\sigma$ \cite{Bikondoa2017, Brown1997}:
\begin{equation}\label{eq:TTC_std}
 G(\vec{q},t_1,t_2)=\frac{\overline{I(t_1)I(t_2)}-\overline{I(t_1)}\cdot\overline{I(t_2)}}{[\overline{I^2(t_1)}-\overline{I(t_1)}^2]^{\frac{1}{2}}\cdot[\overline{I^2(t_2)}-\overline{I(t_2)}^2]^{\frac{1}{2}}}=\frac{\overline{I(t_1)I(t_2)}-\overline{I(t_1)}\cdot\overline{I(t_2)}}{\sigma(t_1)\cdot\sigma(t_2)},
\end{equation}
where $\overline{I}=\langle I\rangle$.

If the scattered intensity fluctuates around a stable mean and has a negative exponential distribution (i.e. fully coherent) \cite{Goodman2020,Pusey1989}, it can be shown that the average speckle intensity equals the standard deviation of speckle intensities \cite{Brown1997,Loudon1983}: 
\begin{equation}
\overline{I}=\sigma = \sqrt{\overline{I^2}-\overline{I}^2}.
\end{equation}

Thus, $ G(\vec{q},t_1,t_2)= Corr(\vec{q},t_1,t_2)-1$, and, the use of $G$-TTC and $Corr$-TTC is physically equivalent under these conditions. In the more general case of partial coherence and non-stable mean intensity, the normalized standard deviation $\beta=\sigma^2 /\langle\mathrm{I}\rangle^2=\beta_{source}*\beta_{sample}$ is a product of speckle contrast due to the properties of the X-ray source / experimental setup \cite{Moller_2021} and fluctuations from the non-constant mean scattering intensity from the sample  \cite{1985Goodman,Pusey1975,DAINTY19771}. Using this relation the \cref{eq:TTC_std} can be rewritten as follows:
\begin{equation}\label{eq:TTC_std_contrast}
 G=\frac{\overline{I(t_1)I(t_2)}-\overline{I(t_1)}\cdot\overline{I(t_2)}}{{\sqrt{\beta}\cdot\overline{I(t_1)}}\cdot\sqrt{\beta}\cdot\overline{I(t_2)}}=\frac{1}{\beta}\cdot \frac{\overline{I(t_1)I(t_2)}}{\overline{I(t_1)}\cdot\overline{I(t_2)}}-\frac{1}{\beta}=\frac{1}{\beta}\cdot (Corr -1).
\end{equation}
This equation denotes the general relation between the two conventional ways of calculating the TTC. List of functions used in the manuscript, as well as their relations (including the general scheme for XPCS data analysis), can be found in \cref{sec:si_relations}. It is important to note that in the case of low-intensity statistics, while \cref{eq:TTC_std_contrast} with such a definition of contrast as well as definitions of $Corr$-TTC and $G$-TTC are mathematically true, they are not operational. In such instances, the contrast needs to be determined from the binomial distribution, and alternative data analysis methods, as well as experimental procedures, are employed instead of TTCs to obtain the intermediate scattering function \cite{Roseker2018,Hua2020,Hruszkewycz2012}. These approaches are beyond the scope of the current article.

\subsection{TTC for non-equilibrium system with evolution of the intensity distribution.}

Classical data analysis assumes the extraction of the one-time correlation function, $g_2$, from the TTC by different coordinate systems (see \cref{sec:CS}) \cite{Bikondoa2017}. For the correlation map defined by the \cref{eq:TTC_mean}, the $g_2$-function may be determined via the generalized Siegert relation:
\begin{equation}\label{eq:g2_1exp}
g_{2_{Corr}}(\vec{q}, \Delta t)=1+\beta(\vec{q})|g_1(t, \Delta t)|^2.
\end{equation}

Depending on the sample dynamics, the $g_1$ function may have different forms. Nevertheless, the standard approach which is valid for the majority of the systems is a  Kohlrausch-Williams-Watts relation: $g_1=\exp \left(-(\Delta t/\tau)^{\gamma}\right)$ with the KWW exponent $\gamma$ and the relaxation time of the system $\tau$ \cite{Williams_Watts_1970}. Substituting \cref{eq:g2_1exp} into \cref{eq:TTC_std_contrast}, it is possible to obtain the $g_2$-function for the correlation map $G$-TTC:
\begin{multline}\label{eq:g2_1exp_G}
g_{2_{G}}(\vec{q}, \Delta t)=\frac{1}{\beta}\cdot (g_{2_{Corr}}(\vec{q}, \Delta t) -1)=\frac{1}{\beta}\cdot \left(1+\beta(\vec{q})|g_1(t, \Delta t)|^2 -1\right)= \\ = |g_1(t, \Delta t)|^2=\exp \left(-2(\Delta t/\tau)^{\gamma}\right).
\end{multline}

Thus, we can conclude that if under the investigated conditions, the Siegert relation is applicable (\cref{eq:g2_1exp}), $g_{2_{G}}$ does not depend on the contrast $\beta$ and is a function of the parameters of dynamics of the system only. This feature will be later used in \cref{sec:CS}. It is important to note that while $g_{2_{G}}$ continues to capture all physical parameters accessible by $g_{2_{Corr}}$, the normalization already performed by the contrast should be taken into account when extracting the parameters of dynamics that depend on it (e.g., non-ergodicity parameter).

The above derivations lead to an important point. The dynamics of a system in equilibrium can be described using the fluctuations of the static structure factor, where the average density is constant, and density fluctuates around this mean value with time. In this equilibrium case, $\overline{I(q)}$ is constant, and dynamics can be directly extracted. Thus, there is no physical difference between the use of $G$-TTC or $Corr$-TTC.

 \begin{figure}
 \centering
 \includegraphics[width=\textwidth]{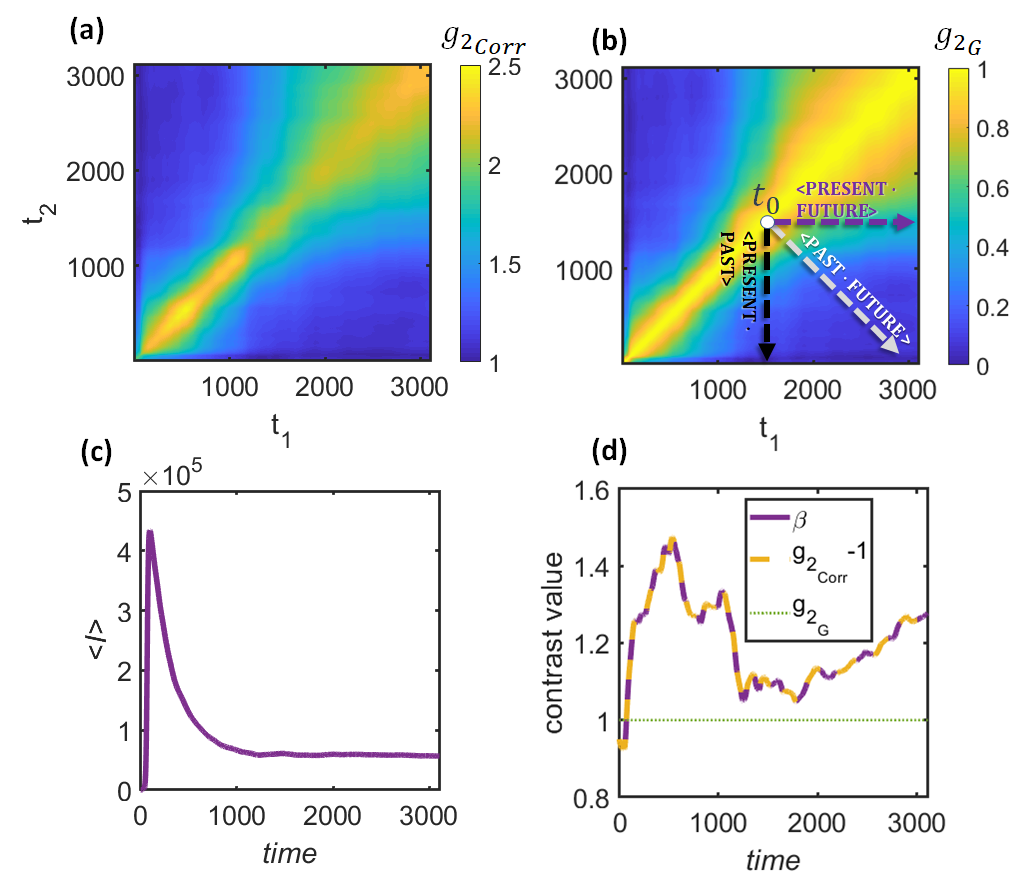}
    \caption{Examples of $Corr$-TTC (a) and $G$-TTC (b) based on Cahn-Hilliard simulations for the same $q$-value. The different options for the $g_2$ - cuts are also illustrated in (b): CCS (purple dashed line) represents the correlation between the present (at time $t_0$) and future dynamics, ACS (grey dashed line) - between the past and future. The black dashed line represents another possibility - the correlation between the present and past. (c) is the evolution of the mean intensity $\overline{I}=\langle I\rangle$ of the system. (d) shows the comparison between the $\beta=\sigma^2/\langle I\rangle^2$ (purple), $g_{2_{Corr}}(t_1=t_2)-1$ (orange) and $g_{2_{G}}(t_1=t_2)$ (green). It can be seen, that $G$-TTC is less sensitive to the kinetic changes of the system than $Corr$-TTC. }
 \label{fgr:example}
\end{figure}

If the system is non-equilibrium and the fluctuations from the sample cannot be described by zero-mean Gaussian statistics, the contrast $\beta$ may evolve with time. Consider a system that exhibits not only dynamics but also kinetic evolution, e.g. a system undergoing phase separation. In contrast to equilibrium systems, there is a change in the behavior of the mean intensity $\overline{I(q)}$ due to its kinetic evolution. This change may lead to a variation of $\beta$. Therefore, for kinetically evolving systems, the calculation of TTCs with $Corr$ (\cref{eq:TTC_mean}) followed by the extraction of $g_{2_{Corr}}$ (\cref{eq:g2_1exp}) may lead to ambiguous data analysis, while the correlation map $G$-TTC (\cref{eq:TTC_std_contrast}) followed by extraction of $g_{2_{G}}$ (\cref{eq:g2_1exp_G}) is not influenced by the contrast evolution.

In order to illustrate the effects for a kinetically evolving system, \cref{fgr:example} shows the example of simulations based on the Cahn-Hilliard equation for spinodal decomposition \cite{Ragulskaya:ti5024, Girelli2021, Cahn1958, Cahn1959} under fully coherent light ($\beta_{source}=1$). The system evolves kinetically, resulting in a significant variation of the mean intensity \textbf{(c)}. The $G$-TTC is stable, while the $Corr$-TTC shows fluctuations along the diagonal (compare \textbf{(a)} and \textbf{(b)}) with values that sometimes exceed 2 due to the limited number of scatterers \cite{DAINTY19771, Pusey1975}. These fluctuations are the same as the contrast of the system, $\beta=\sigma^2/\langle I\rangle^2=\beta_{source}*\beta_{sample}=\beta_{sample}$, and reflect the change in the scattering intensity distribution \textbf{(d)}.

\section{Analysis of two-time correlation functions using different time coordinate systems}\label{sec:CS}
\subsection{Quantitative analysis of TTCs}

Quantitative description of the evolution of correlation function typically requires slicing ("cutting") the TTC at distinct observation times $t_0$ and extracting the parameters of dynamics of the system, such as relaxation time, $\tau(t_0)$, and Kohlrausch-Williams-Watt exponent, $\gamma(t_0)$.

The evolution of the relaxation time of a non-equilibrium system is frequently used for the calculation of the evolution of the diffusion coefficient, velocity, or other macroscopic observables of the system \cite{Chushkin2022,Reiser2022-pj,Czakkel_2011,Lehmkhler2020}. Through this approach, a time series of cuts and corresponding parameters ($\tau(t_0)$, $\gamma(t_0)$) obtained from the TTC results in a time-resolved evolution of macroscopic observables of the system under study. 

The foundation of this quantitative analysis is based on derivations made for \textit{equilibrium} systems. Therefore, such treatment of the experimental data requires that $\tau(t_0)$ represents an "instantaneous" description of the system, i.e. it is obtained for a specific moment in time $t_0$ \cite{Ladd1995}. We will refer to such dynamics as effective dynamics. For instance, $\tau(t_0)$ is considered not to be affected by any possible future ($t>t_0)$ perturbations of the system (e.g. beam damage) \cite{Ruta2017b, Reiser2022-pj,Chushkin2022,Timmermann2023}. This can be interpreted as the relaxation time of the corresponding equilibrium system, wherein the macro-parameters align with the investigated moment of the non-equilibrium system. 

To this end, one needs to estimate the relaxation time of the corresponding equilibrium system based on TTC from a non-equilibrium system. It is typically done by extracting $g_2$ functions in the form of one-dimensional cuts. There are several ways to obtain cuts from TTCs that are discussed below.

\subsection{Frequently employed coordinate systems.}\label{sec:coordinate}

The most frequently employed methods for extracting $g_2$ functions are horizontal and diagonal cuts as illustrated in \cref{fgr:example} (b).

The diagonal cuts were introduced alongside TTCs by Brown \textit{et al.} \cite{Brown1997} in 1997. In the context of X-ray photon correlation spectroscopy (XPCS), diagonal cuts have a longstanding history and were commonly used in the past \cite{Malik1998, Brown1999, Livet2001, Sutton2003, Fluerasu2005, Muller2011, Orsi2010, Bikondoa_2012, Ruta2012}. In 2017, Bikondoa presented horizontal cuts as an alternative to diagonal cuts \cite{Bikondoa2017}. Since the horizontal cuts were argued to be more intuitive and consistent with the standard calculations in statistical mechanics, they were termed ``conventional coordinate system'' (CCS).  Moreover, it was discussed that the use of diagonal cuts could lead to interpretation issues when external forces or perturbations are present (e.g., such as the system presented in \cite{Ruta2017b}) since it might be argued that they mix events prior and subsequent to the perturbation. Consequently, the diagonal cuts were referred to as the ``alternative coordinate system'' (ACS) according to \cite{Bikondoa2017}. As a result, the CCS approach has gained popularity and is increasingly used \cite{Zhang2021, Girelli2021, Lehmkuhler2021}. Nonetheless, the traditional ACS approach is still being used today, inter alia, for facilitating comparison with earlier XPCS studies.

As the name suggests, the diagonal cuts $g_{2_{ACS}}(t_0, \Delta t)$ at different sample age $t_0=t_{age}=(t_1+t_2)/2$ are obtained by taking the line perpendicular to the $t_1=t_2$ diagonal with the delay time  $\Delta t = |t_2-t_1|$. The horizontal cuts at different waiting times $t_0=t_w$ are performed by extraction of line with $t_1=constant$ (or $t_2=constant$) and delay time defined as $\Delta t :=\left|t_{2}-t_{1}\right|$. In this manner, the diagonal cuts (ACS) represent correlations between the past and the future states of the system relative to the considered moment $t_0$, whereas the horizontal cuts represent correlations of the system at moment $t_0$ with its future states. Therefore, in fact, the horizontal cuts may as well mix events prior and subsequent to the perturbation if the latter happens in the future. We note that for the complete description, one may also consider the vertical cuts (see \cref{fgr:example} (b)), i.e. correlations of the system at moment $t_0$ with its past states. In the following, we mainly focus on ACS and CCS, but this third option can be straightforwardly derived similarly to CCS.

Importantly, for non-equilibrium systems, the ACS and CCS can produce different evolutions of $\tau(t_0)$ and $\gamma(t_0)$. These values play a vital role in interpreting the results, and as such, their correct determination is crucial for characterizing the dynamics of the system.

\subsection{Connection between ACS and CCS for equilibrium and non-equilibrium systems.}

First, we discuss how the ACS and CCS correlation functions are connected. They can be defined as:

\begin{equation}\label{eq:g1ccs}
    g_{1_{CCS}}(t, \Delta t) = \langle E^*(t) E(t + \Delta t) \rangle_{N}
\end{equation}

and

\begin{equation}\label{eq:g1acs}
     g_{1_{ACS}}(t, \Delta t) = \langle E^*(t - \Delta t / 2) E(t + \Delta t / 2) \rangle_{N},
\end{equation}
where $ \langle\rangle_{N}$ is the average over the ensemble.

 We note that for each definition of the correlation function, the Siegert relation holds (\cref{eq:g2_1exp}):

\begin{equation}
    g_{2_{Corr, CCS}}(t, \Delta t) = 1 + \beta|g_{1_{CCS}}(t, \Delta t)|^2,
\end{equation}

\begin{equation}
     g_{2_{Corr, ACS}}(t, \Delta t) = 1 + \beta|g_{1_{ACS}}(t, \Delta t)|^2.
\end{equation}

In the case of equilibrium systems, correlation functions are translation invariant in time, i.e.:

\begin{equation}
    g_{1_{CCS}}(t, \Delta t) = g_{1_{CCS}}(t + T, \Delta t) = g_{1_{CCS}}(0, \Delta t),
\end{equation}

\begin{equation}
    g_{1_{ACS}}(t, \Delta t) =  g_{1_{ACS}}(t + T, \Delta t) =  g_{1_{ACS}}(0, \Delta t),
\end{equation}
so they only depend on $\Delta t$. Therefore, $ g_{1_{CCS}}$ and $ g_{1_{ACS}}$ are equivalent in the case of equilibrium systems, since 

\begin{equation}\label{eq:acs_ccs_erg}
    g_{1_{ACS}}(t, \Delta t) =  g_{1_{ACS}}(t + \Delta t/ 2, \Delta t) =  g_{1_{CCS}}(t, \Delta t).
\end{equation}

For non-equilibrium systems, this is not the case, as the first equality in \cref{eq:acs_ccs_erg} generally does not hold. However, the second equality still holds by definition, so that for non-equilibrium we expect generally:
\begin{equation}\label{eq:g1_non-equilibrium}
     g_{1_{ACS}}(t, \Delta t) \neq  g_{1_{ACS}}(t + \Delta t/ 2, \Delta t) =  g_{1_{CCS}}(t, \Delta t).
\end{equation}
This relation only reveals a trivial transformation of the variables. Nevertheless, it shows how $ g_{1_{CCS}}$ and $ g_{1_{ACS}}$ are related in non-equilibrium systems, and that, in general, they are not identical. 

\subsection{Geometrical illustration of ACS and CCS.}

 \begin{figure}
 \centering
 \includegraphics[width=\textwidth]{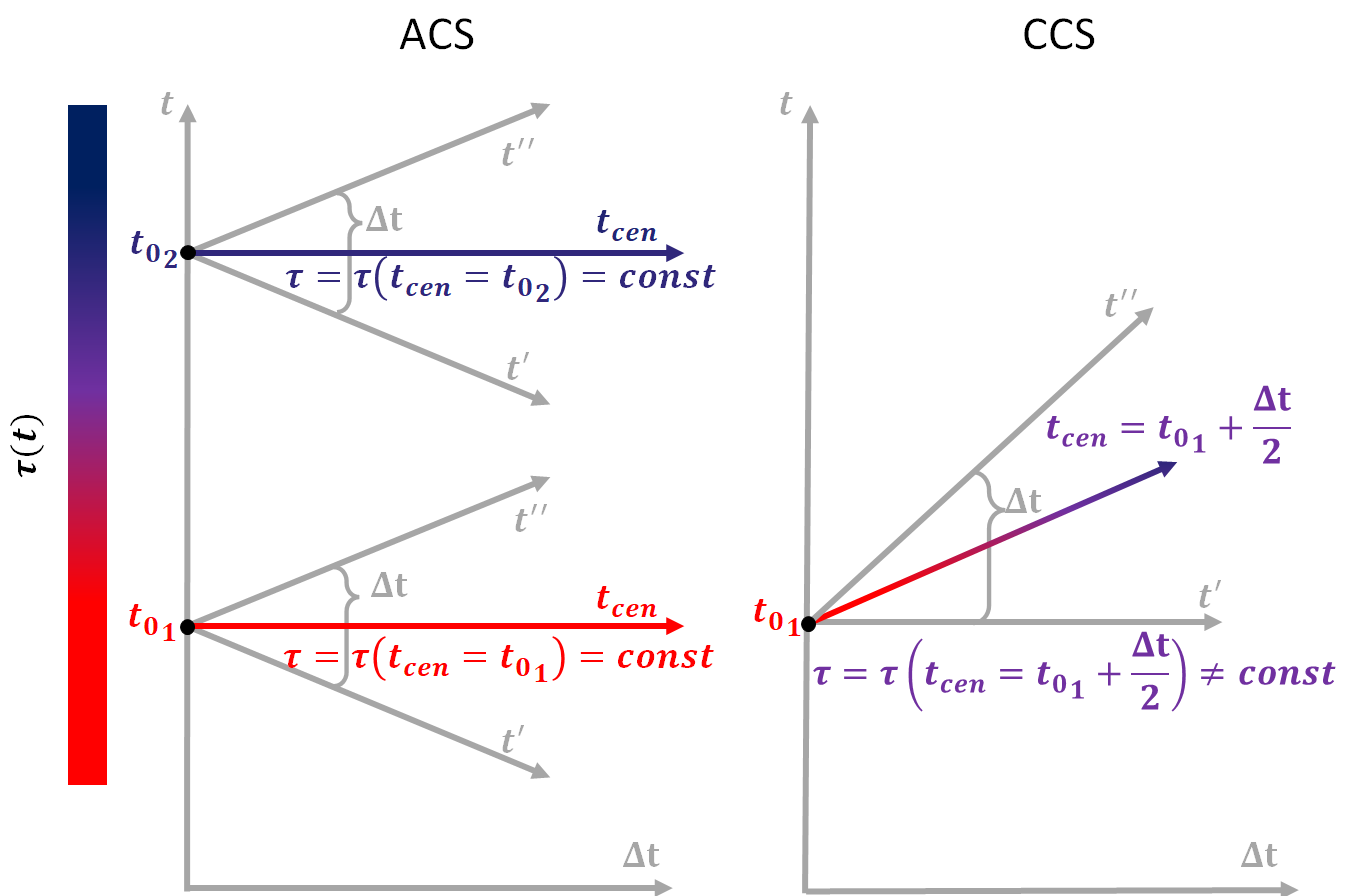}
    \caption{Schematic description of obtaining dynamics in a non-equilibrium system through ACS (left panel) and CCS (right panel) in the ($t$, $\Delta t$) plane ($t'=t'(t, \Delta t)$, $t''=t''(t, \Delta t)$). The time dependency is illustrated by different colors, gradually changing from red to deep blue. For ACS: $t'=t-\Delta t/2$ and $t''=t+\Delta t/2$. Therefore, at any time moment $t_0$: $t_{cen}=(t'+t'')/2=t_0$, and, thus, calculated $\tau$ can be approximated by the effective dynamics with $\tau(t_0)$ (see examples for $t_{0_{1}}$ and $t_{0_{2}}$). For CCS: $t'=t$ and $t''= t+\Delta t$. Therefore, at any time moment $t_0$: $t_{cen}=(t'+t'')/2=t_0+\Delta t/2$. Thus, the relaxation time $\tau=\tau(t_0 + \Delta t/2)$ cannot be approximated by effective dynamics anymore, except when the system is quasi-equilibrium and $\tau(t_0 + \Delta t/2) \approx \tau(t_0)$. }
 \label{fgr:cut_scheme}
\end{figure}

The disparity between ACS and CCS extends beyond mere mathematical expressions. Instead, it lies in their fundamental approach to correlation. CCS correlates the system with itself in the future, whereas ACS reveals the correlation between the system in the future and itself in the past, with equal distance from the time under investigation. These approaches are illustrated in \cref{fgr:cut_scheme}. As was discussed before,  it is assumed that the correlation function $\langle E^*(t') E(t'') \rangle_{N}$ between times $t'$ and $t''$ is equal to the correlation function of the corresponding \textit{equilibrium system} with some fixed dynamical property $\tau_{\mathrm{instantaneous}} = \tau(t_{cen})$. Furthermore, it is natural to assume that the time moment $t_{cen}$ is the average $t_{cen} = (t' + t'') / 2$ (see \cref{fgr:cut_scheme}). If this is the case, at any moment $t_0$ in time, we shall consider the ACS definition of the correlation function (diagonal cut) as the one that more closely resembles the effective dynamics of the system $\tau(t_0)$, while the horizontal cut will correspond to $\tau(t_0 + \Delta t / 2)$. Therefore, the $g_{2_G}$-functions for ACS and CCS can be represented as follows:

\begin{equation}\label{eq:ACS_Theory}
g_{2_{G, ACS}}(t_0, \Delta t)=\exp \left(-2\left(\frac{\Delta t}{\tau\left(t_0\right)}\right)^{\gamma\left(t_0\right)}\right),
\end{equation}
\begin{equation}\label{eq:CCS_Theory}
g_{2_{G, CCS}}(t_0, \Delta t)=\exp \left(-2\left(\frac{\Delta t}{\tau\left(t_0+\frac{\Delta t}{2}\right)}\right)^{\gamma\left(t_0+\frac{\Delta t}{2}\right)}\right).
\end{equation}

Obviously, for an equilibrium process, $\tau(t_0)=\tau (t_0 + \Delta t /2) = const$ and $\gamma(t_0)=\gamma (t_0 + \Delta t /2) = const$, which leads to the same results for ACS and CCS. However, in the case of a non-equilibrium process, only the ACS has the conventional form of the $g_2$ - function, which is typically used for data analysis for the extraction of the relaxation time $\tau (t_0)$: $g_{2_{G, ACS}}(t_0, \Delta t)=g_{2_{G}}(t_0, \Delta t)$  (compare \cref{eq:ACS_Theory} and \cref{eq:g2_1exp_G}). In contrast for each $\Delta t$, $g_{2_{CC S}}$ represents a different set of $\tau(t_0 + \Delta t/2)$ and $\gamma (t_0 + \Delta t /2)$ pair. Therefore, while the ACS cuts may be used to describe the effective dynamics, followed by extraction of the momentary properties of the system, the CCS cuts provide some averaged properties of the correlation between the considered moment and the future. These fundamental differences should be taken into account for the interpretation of the results of the XPCS data analysis. 

We note that while the ACS may correspond to the dynamics of the corresponding equilibrium of the non-equilibrium system, it is not necessarily guaranteed. In the following, we demonstrate the system for which this assumption holds true.

\subsection{Simulation example}

In this section, we demonstrate the use case of ACS cuts for the extraction of effective dynamics. We employ the simulations for a model example of a non-equilibrium system and compare ACS and CCS with the corresponding equilibrium for each sample age. 

The model system consists of a set of particles. The position of a particle is represented by the vector $\textbf{r}=(x,y)$, for which the movement along the Cartesian coordinates $x$ and $y$ is statistically independent. The probability density function $P(\textbf{r},t) = P(x,t)P(y,t)$ is set to follow:
\begin{equation}\label{eq:model}
    \begin{aligned}
    P(x, t)=\frac{1}{\sqrt{4 \pi D(t) t}} \exp \left(-\frac{\left(x-x_0\right)^2}{4 D(t) t}\right),
    \\P(y,t)=\frac{1}{\sqrt{4 \pi D(t) t}} \exp \left(-\frac{\left(y-y_0\right)^2}{4 D(t) t}\right),
    \end{aligned}
\end{equation}
where $t$ is time, $x_0$ and $y_0$ are initial positions of the particle. At each discrete time (sample age) of simulated observation, we modify the $D(t)$ parameter of the system to simulate non-equilibrium dynamics. This approach allows us to obtain both the TTC of the non-equilibrium system and the corresponding equilibrium parameters for each discrete moment $t_0$. Parameters of simulations can be found in SI \cref{sec:SI_parameters}.

While, in the general case, such a non-equilibrium system does not correspond to any known example of anomalous diffusion, and the corresponding equation of motion remains unclear, for each specific $t_0$, \cref{eq:model} describes "momentarily" Brownian motion, which serves as a corresponding equilibrium scenario of this non-equilibrium system. In this case, $D(t_0)$ is the diffusion coefficient of this corresponding equilibrium and the relaxation time $\tau(t_0)_{cor. eq.}\propto 1/D(t_0)$. Furthermore, $\gamma(t_0)_{cor. eq.}$ is equal to $1$ and $\Gamma(t_0)_{cor. eq.}=1/\tau(t_0) \propto q^2$. The values of $\tau(t_0)_{cor. eq.}$ and $\gamma(t_0)_{cor. eq.}$ parameters for the equilibrium were double-checked by performing a simulation with constant $D=D(t_0)$.

 \begin{figure}
 \centering
 \includegraphics[width=\textwidth]{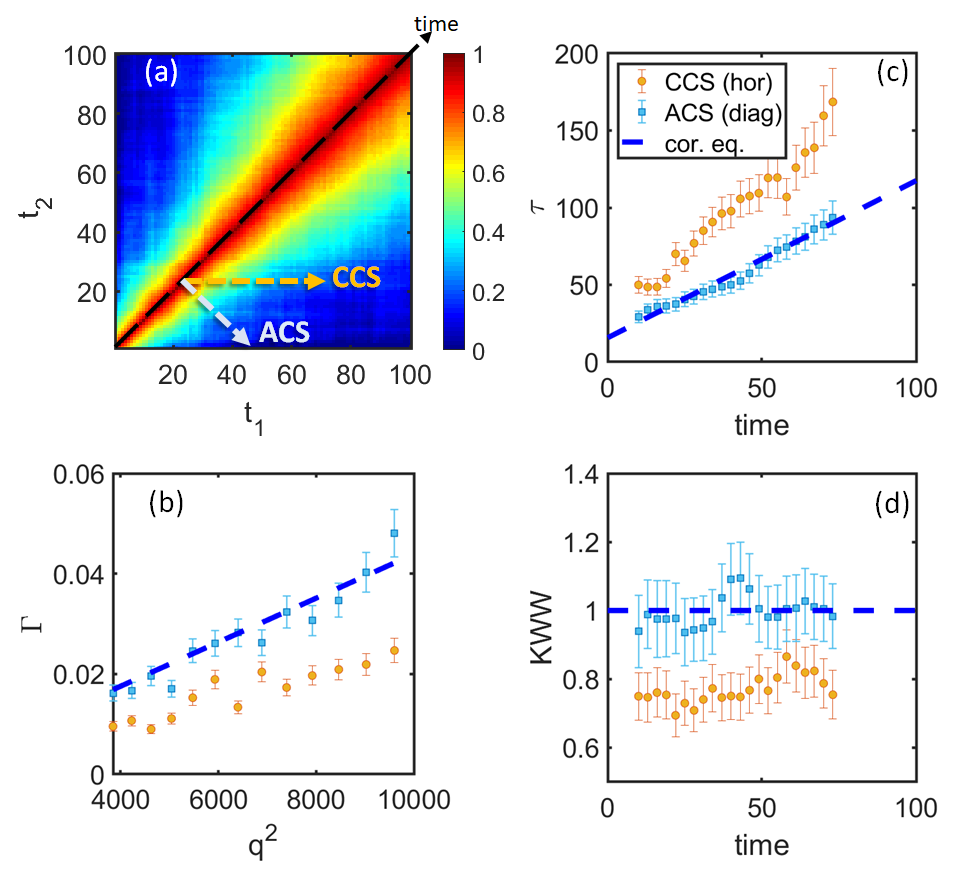}
    \caption{Data analysis for a model system for \textit{Case 1a} -  a linear increase in $1/D$. (a) $G$-TTC for $q=74$ pixel. (b) Relaxation rate $\Gamma$ as a function of $q^2$ at $time = 25$. (c) and (d) represent relaxation time $\tau$ and $\gamma$ as functions of time, correspondingly. Orange circles display results for CCS analysis, light blue squares - for ACS, and dashed blue line shows results from corresponding equilibrium systems. Results for other $q$ and $time$ values are similar. }
 \label{fgr:cuts}
\end{figure}

For the sake of illustration, we demonstrate three cases of the evolution of $1/D$=$1/D(t)$ parameter: linear (\textit{Case 1}), exponential (\textit{Case 2}), and sinusoidal (\textit{Case 3}). In the \textit{Case 1}, we assume that $1/D$ of the model system were to change linearly for each discrete time step $j$:
\begin{equation}\label{eq:viscosity}
1/D(j+1) = constant + 1/D(j), \; j=[1,N_t] \text{,}
\end{equation}
where $N_t$ is the total number of time steps of the simulation. In this case, a linear change in the relaxation time of the corresponding equilibrium scenario $\tau(t_0)_{cor. eq.}=\tau(j)_{cor. eq.}\propto 1/D(j)$ with time is expected. Depending on the sign of the $constant$ in \cref{eq:viscosity}, the system slows down (\textit{Case 1a}) or accelerates (\textit{Case 1b}). 

\begin{figure}
 \centering
 \includegraphics[width=\textwidth]{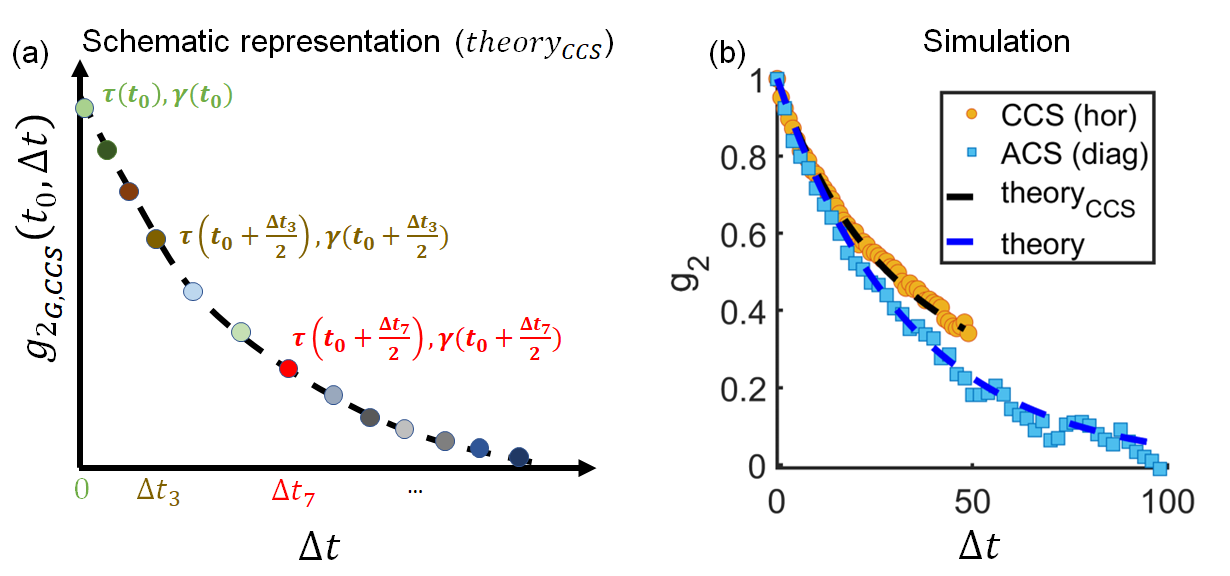}\caption{(a) Schematic representation of $g_{2_{G, CCS}}$, which follows \cref{eq:CCS_Theory} and \cref{fgr:cut_scheme}. For each $\Delta t$,  $g_{2_{G, CCS}}$ represents a different set of $\tau(t_0+\Delta t/2)$ and $\gamma(t_0+\Delta t/2)$
pair. (b) Comparison of $g_2$-cuts for the simulated system for \textit{Case 1a}, presented in \cref{fgr:cuts} for $q=74$ pixels and $time = 50$. The CCS and ACS cuts from the simulated TTC are presented with orange circles and blue squares, respectively. The black dash line shows the $g_{2_{G, CCS}}$, calculated via \cref{eq:CCS_Theory}. Blue dash line shows conventional $g_{2_{G}}$ function, calculated via \cref{eq:g2_1exp_G}.}
\label{fgr:g2_cuts}
\end{figure}

The results of the \textit{Case 1a} simulation of a  linear slowdown ($constant>0$) are presented in \cref{fgr:cuts}. Both, CCS and ACS qualitatively capture the linear behavior of the relaxation time as well as $\Gamma(t)=1/\tau(t) \propto q^2$ behavior. Furthermore, in this case, the ACS manages to capture the momentary description and, thus, the effective dynamics. Remarkably, the relaxation time as well as $\gamma$ have values are similar to the corresponding equilibrium equivalents. On the other hand, CCS does not provide a momentary description of the system. Instead, the obtained relaxation time for all $q$-values is larger than for the corresponding equilibrium scenarios and also suggests a different slope (see \cref{fgr:cuts} (b) and (c)). Furthermore, the $\gamma$ is around 0.75 (\cref{fgr:cuts} (d)) in comparison to 1 for the equilibrium system. The CCS results come as a direct reflection of the slowdown of the system with time. As was discussed earlier, the $g_{2_{G, CC S}}$ for each $\Delta t$ represents a different set of $\tau(t_0 + \Delta t/2)$ and $\gamma (t_0 + \Delta t /2)$ pair (see schematic representation in \cref{fgr:g2_cuts} (a) and \cref{eq:CCS_Theory}).  For our model system, for any positive $\Delta t$: $\tau(t_0)< \tau (t_0 + \Delta t/2)$ and $\gamma(t)_{cor. eq.}=1$, therefore, $g_{2_{G, CCS}}$ is stretched in comparison to $g_{2_{G, ACS}}$ (see \cref{eq:CCS_Theory} and \cref{eq:ACS_Theory}). Similarly, if this system is accelerated (\textit{Case 1b}), the $g_{2_{G, ACS}}$ will catch the effective dynamics, while  $g_{2_{G, CCS}}$ will be compressed in comparison to $g_{2_{G, ACS}}$. This conclusion is demonstrated in the simulation of \textit{Case 1b} in SI \cref{fgr:cuts_lin_decrease}.

This conclusion can be further supported by the comparison of the ACS and CCS $g_2$-cuts of our model-system with the conventional $g_2$-function (\cref{eq:g2_1exp_G}) and $g_{2_{CCS}}$ (\cref{eq:CCS_Theory}), both calculated from instantaneous $\tau$, obtained from corresponding equilibrium equivalents and presented in \cref{fgr:g2_cuts} (b). The CCS $g_2$-cuts align well with the estimation via \cref{eq:CCS_Theory} and ACS $g_2$-cuts overlap with the conventional representation (\cref{eq:g2_1exp_G}). 
\begin{figure}
 \centering
 \includegraphics[width=\textwidth]{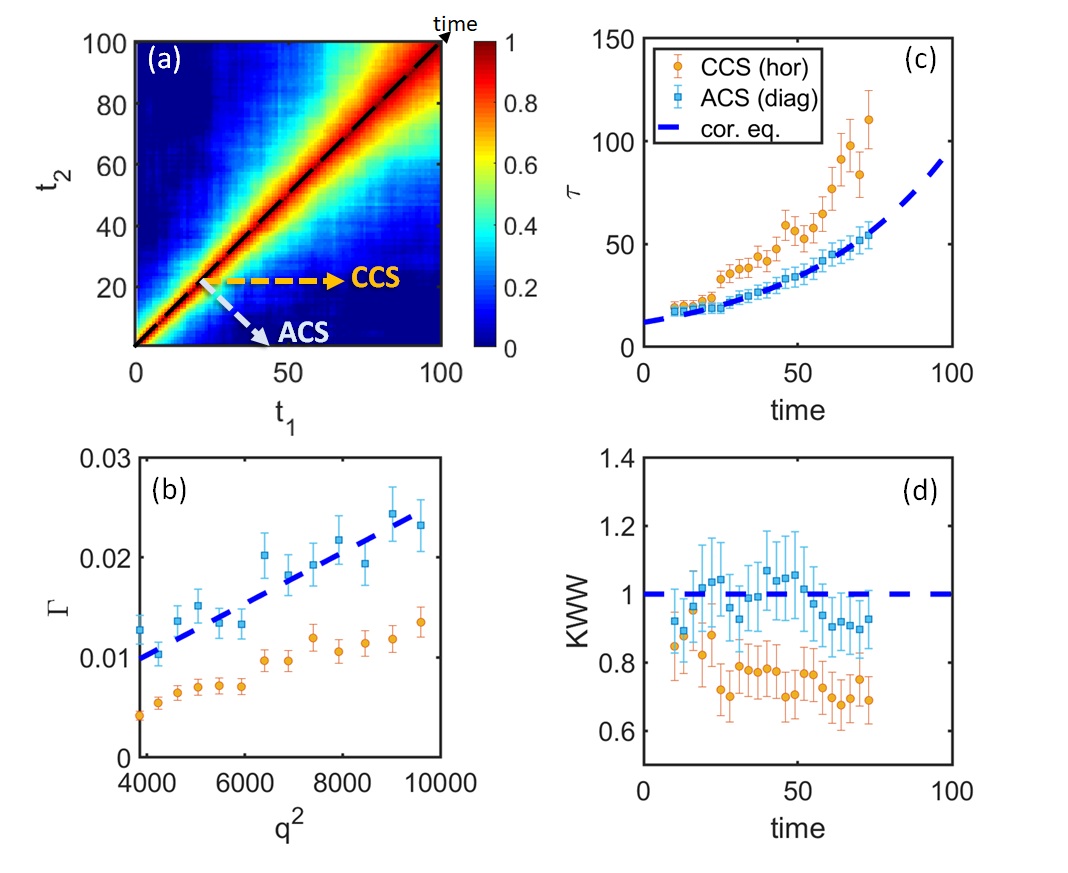}
    \caption{Data analysis for a model system for \textit{Case 2} - an exponential increase in $1/D$ (to be compared to \cref{fgr:cuts} (linear increase)). (a) $G$-TTC for $q=86$ pixels. (b) Relaxation rate $\Gamma$ as a function of $q^2$ at $time = 70$. (c) and (d) represent relaxation time $\tau$ and $\gamma$ as functions of time, correspondingly. Orange circles display results for CCS analysis, light blue squares - for ACS, and dashed blue line shows theoretical behavior. Results for other $q$ and $time$ values are similar. }
 \label{fgr:cuts_exp}
\end{figure}

 \begin{figure}
 \centering
 \includegraphics[width=\textwidth]{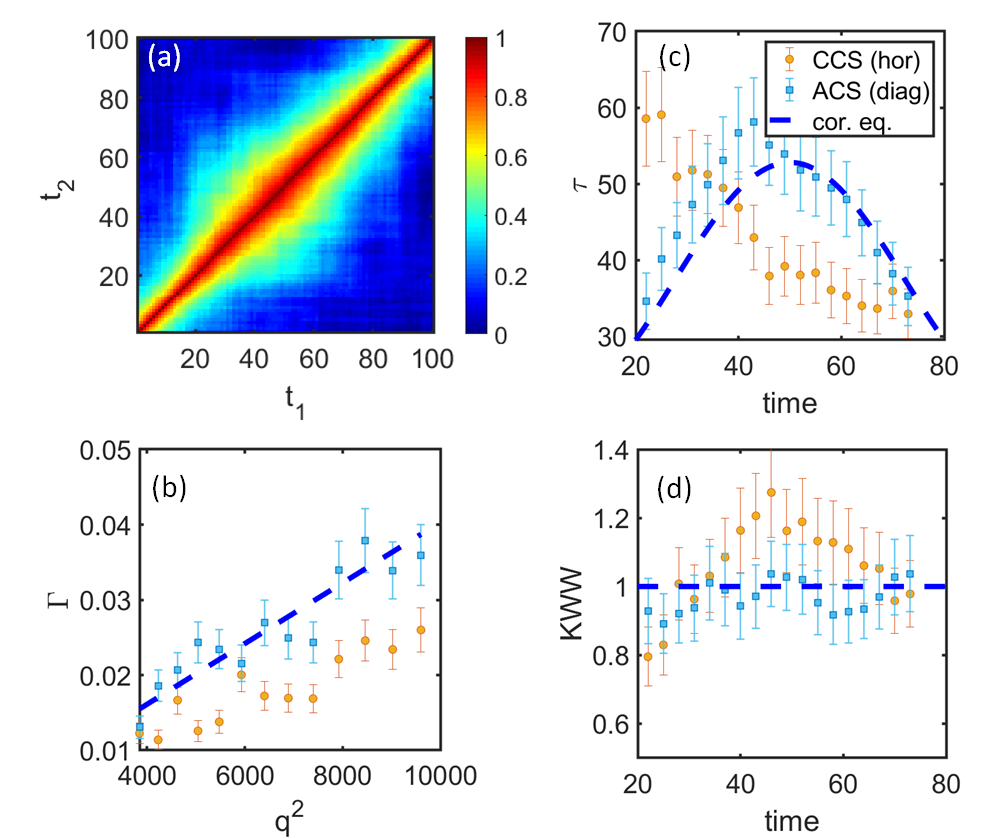}
    \caption{Data analysis for a model system for \textit{Case 3} - a sinusoidal change in $1/D$ (to be compared with \cref{fgr:cuts} and \cref{fgr:cuts_exp} (linear and exponential case)). (a) $G$-TTC for $q=83$ pixel. (b) Relaxation rate $\Gamma$ as a function of $q^2$ at $time = 25$. (c) and (d) represent relaxation time $\tau$ and $\gamma$ as functions of time, correspondingly. Orange circles display results for CCS analysis, light blue squares - for ACS, and dashed blue line shows results from corresponding equilibrium systems. Results for other $q$ and $time$ values are similar. }
 \label{fgr:cuts_sin}
\end{figure}

These general conclusions also remain valid for a possible nonlinear evolution of the $1/D$.  \cref{fgr:cuts_exp} (\textit{Case 2}) and \cref{fgr:cuts_sin} (\textit{Case 3}) demonstrate exponential and sinusoidal change in $1/D$ of the investigated system, correspondingly. In both cases, ACS data analysis results overlap with the corresponding equilibrium scenario and capture the effective dynamics, while CCS results represent some averaged dynamics and reflect the acceleration or slowdown of the system. For example, $\gamma$ and $\tau$ in the \textit{Case 3} correspond to $\gamma_{CCS}<\gamma_{cor. eq.}=1$ and $\tau_{CCS}>\tau_{cor. eq.}$ in the accent of the sinus, while $\gamma_{CCS}>1$ and $\tau_{CCS}<\tau_{cor. eq.}$ for the descent part of the sinus, which is consistent with the previous discussions.

Therefore, ACS and CCS analysis of the system give generally different descriptions with different quantities of key parameters which leads to a different qualitative interpretation of the dynamics. The ACS analysis of various simulations of the model system suggests that the system under investigation can be described by Brownian motion momentarily at each measurement time, capturing the effective dynamics. The CCS analysis suggests subdiffusion or superdiffusion behavior with a relaxation time evolution distinct from the corresponding equilibrium scenario. The observed discrepancies arise as a natural outcome of the functional form of the corresponding $g_2$-functions. Nevertheless, the prevailing XPCS data analysis does not encompass an assessment of the distinctions between the two types of cuts and their resulting quantitative effects on the relaxation time and KWW parameters, which may lead to ambiguous interpretation. 

\section{Summary and conclusions}
In this article, we discussed the analysis of XPCS data, focusing specifically on the comparison of two widely used TTC calculation methods: normalization by the mean ($Corr$-TTC) and normalization by the standard deviation ($G$-TTC). We  demonstrate that for kinetically evolving systems, $Corr$-TTC is susceptible to intensity variations, potentially leading to inconclusive data interpretation. In contrast, $G$-TTC is generally robust against these fluctuations. Therefore, we recommend using $G$-TTC for analyzing processes such as film growth, coarsening, phase separation and etc. 

We then compared the two widely used methods for extracting one-time correlation and relevant information from the TTC: the ACS introduced by Brown \cite{Brown1997} and CCS  recommended by Bikondoa \cite{Bikondoa2017}. While for equilibrium systems these methods produce consistent results, for non-equilibrium systems ACS and CCS can produce distinct evolutions of the relaxation time ($\tau$) and Kohlrausch-Williams-Watt exponent ($\gamma$), which are crucial for interpreting experimental data. 

Based on a geometrical representation, we derived the functional forms of $g_2$ for ACS and CCS cuts. Our analysis revealed that for non-equilibrium systems, ACS yields the conventional (theoretical) form of the $g_2$ - function, which is typically used for data analysis to extract the relaxation time $\tau (t_0)$: $g_{2_{G, ACS}}(t_0, \Delta t)=g_{2_{G}}(t_0, \Delta t)$. In contrast, the $g_{2_{CC S}}$ for each $\Delta t$ corresponds to a distinct set of $\tau(t_0 + \Delta t/2)$ and $\gamma (t_0 + \Delta t /2)$ pairs. We demonstrated these dependencies through simulations of case studies. Therefore, while ACS cuts may be used to extract the effective properties of the system, CCS reflects the average properties of  correlations between the considered time point and the future.  These distinctions between the two analysis methods and their consequential quantitative impacts on the relaxation time and KWW parameters should be carefully considered when interpreting XPCS data. For instance, if the experimental results are compared to theoretical predictions, it is important to first identify the analysis method that aligns with the assumptions of the theoretical model used. As we are lacking general analytical expressions for non-equilibrium TTC correlation functions, the TTC analysis would greatly benefit from guidance through further simulation work.

\section{Conflicts of interest.}
There are no conflicts to declare.

\section{Acknowledgements.}
This work was supported by the DFG and BMBF
(05K19PS1, 05K20PSA, 05K22PS1, 05K20VTA, 05K22VTA).
The authors would like to thank Martin Oettel for productive discussions.

\section{Supporting information}

\subsection{Parameters of simulation}\label{sec:SI_parameters}
The simulation is performed on a 2D grid 512 × 512 for 100 time steps with $\Delta t = 1$ for a set of 512 non-touching particles. The boundary conditions and the particle-particle interactions follow the implementation in \cite{Ragulskaya:ti5024}. Specifically, we avoid interactions with boundaries and other particles by setting appropriate size parameters and initial particle coordinates. In rare cases when such interactions occur, the simulation is terminated and repeated with other initial positions of particles. For each particle the probability density function was defined via \cref{eq:model}. The speckle pattern (i.e. image in reciprocal space $I(\mathbf{q},t_{age})$) can be calculated for each time step as a square of the magnitude of the 2D fast Fourier transform of the fluctuations of the concentration \cite{Barton}. The outcome is similar to the 2D scattering pattern obtained via XPCS experiments. This pattern was then used to calculate the TTC for the simulation via \cref{eq:TTC_std}. 

\subsection{Functions and their relations}\label{sec:si_relations}

\begin{itemize}
    \item \textbf{"\textit{Corr}-TTC"} - calculation of the TTC via $Corr$ (see \cref{eq:TTC_mean}) - the normalization to the mean intensity.
    \item \textbf{"\textit{G}-TTC"} - calculation of the TTC via $G$ (see \cref{eq:TTC_std}) - the normalization to the standard deviation of the intensity.
        
\end{itemize}

\begin{table}
    \centering
    \begin{tabular}{|c||c|c|} \hline 
         & ACS & CCS \\ \hline \hline
        $g_1$ &  \begin{tabular}{@{}c@{}} $g_{1_{ACS}}(t, \Delta t) = \langle E^*(t - \Delta t / 2) E(t + \Delta t / 2) \rangle_{N}$ \\ \cref{eq:g1acs} \end{tabular}  & \begin{tabular}{@{}c@{}}$g_{1_{CCS}}(t, \Delta t) = \langle E^*(t) E(t + \Delta t) \rangle_{N}$ \\ \cref{eq:g1ccs}\end{tabular}  \\ \hline 
    \end{tabular}
    
    \caption{Definition of $g_1$-function for ACS and CCS cuts.}
    \label{tab:g1_def}
\end{table}

\begin{table}
    \centering
    \begin{tabular}{|c||c|c|} \hline 
         & $G$ (\cref{eq:TTC_std}) & $Corr$ (\cref{eq:TTC_mean}) \\ \hline \hline
        $g_2$ & \begin{tabular}{@{}c@{}} $g_{2_{G}}=|g_1(t, \Delta t)|^2$ \\ \cref{eq:g2_1exp_G}\end{tabular} & \begin{tabular}{@{}c@{}}$g_{2_{Corr}}=1+\beta(\vec{q})|g_1(t, \Delta t)|^2$ \\ \cref{eq:g2_1exp}\end{tabular}  \\ \hline 
        $g_{2_{ACS}}$ & $g_{2_{G, ACS}}=|g_{1_{ACS}}(t, \Delta t)|^2$ & $g_{2_{Corr, ACS}}=1+\beta(\vec{q})|g_{1_{ACS}}(t, \Delta t)|^2$ \\ \hline
        $g_{2_{CCS}}$ & $g_{2_{G, CCS}}=|g_{1_{CCS}}(t, \Delta t)|^2$ & $g_{2_{Corr, CCS}}=1+\beta(\vec{q})|g_{1_{CCS}}(t, \Delta t)|^2$ \\ \hline
    \end{tabular}
    
    \caption{Definition of $g_2$-function, extracted from $G$ and $C$-TTCs (first and second column, respectively): in general and for ACS and CCS cuts (first, second and third rows, respectively) in case of valid Siegert relation.}
    \label{tab:g1_def}
\end{table}

 \begin{figure}
 \centering
 \includegraphics[width=\textwidth]{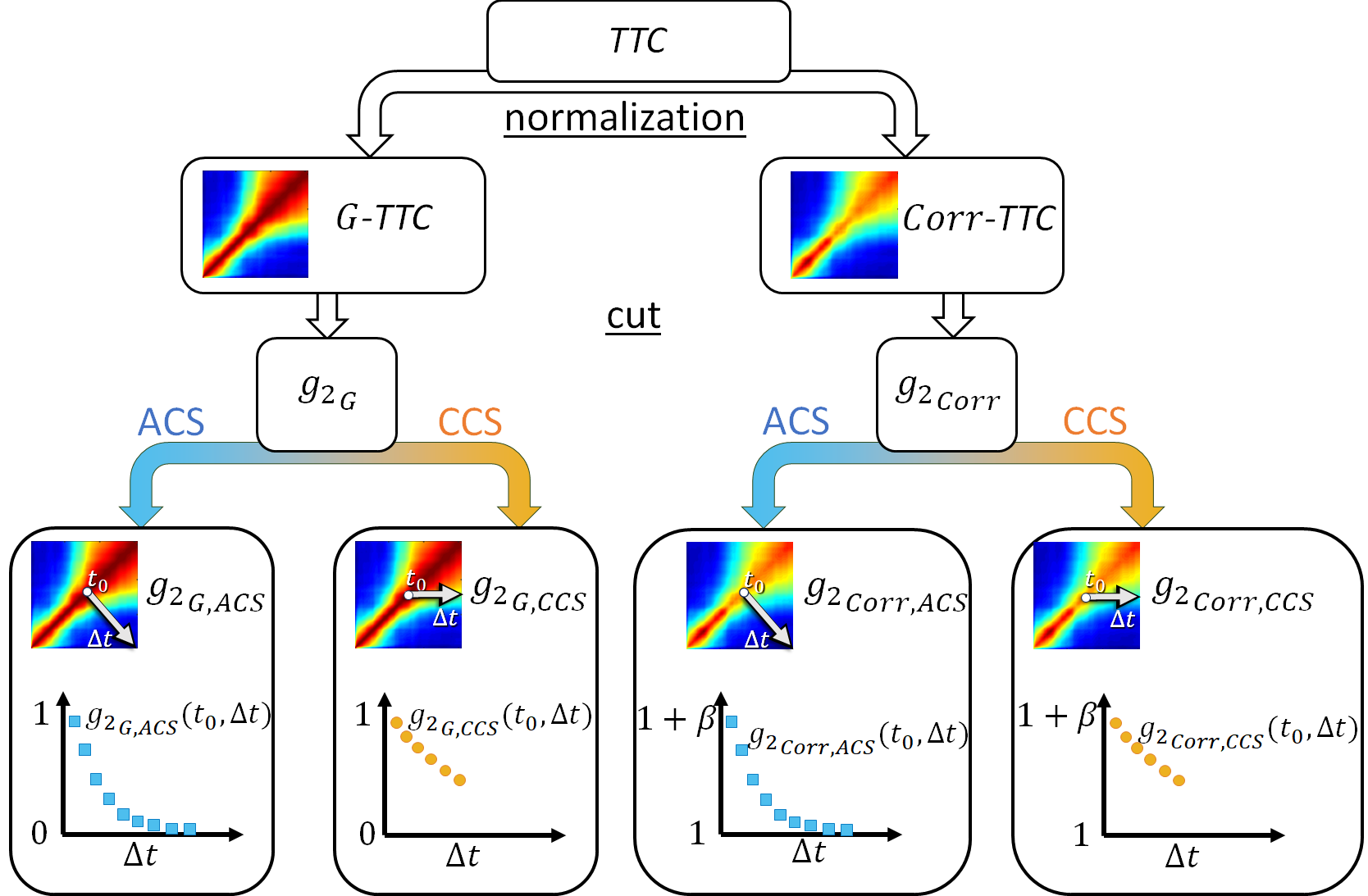}
    \caption{Scheme of relations between different functions, used in XPCS data analysis.}
 \label{fgr:cuts_lin_decrease}
\end{figure}

\subsection{Additional figures}

 \begin{figure}
 \centering
 \includegraphics[width=\textwidth]{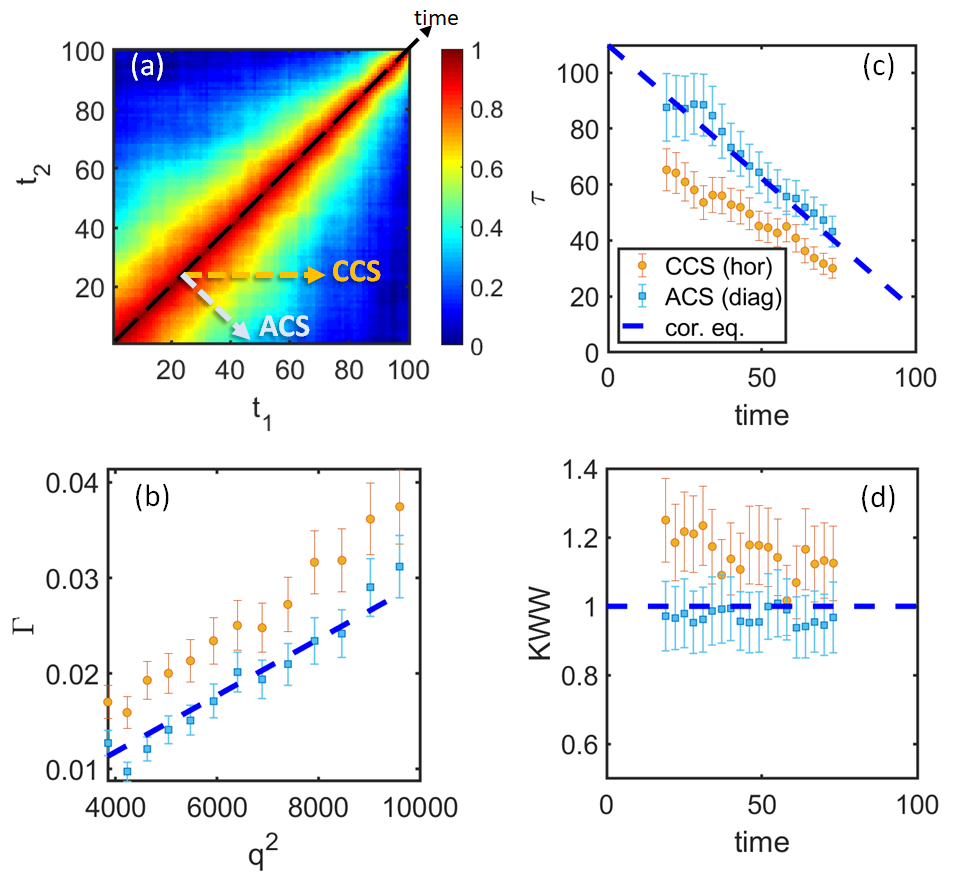}
    \caption{Data analysis for a model system for \textit{Case 1b} - a linear decrease in $1/D$ (to be compared with \cref{fgr:cuts}). (a) $G$-TTC for $q=77$ pixel. (b) Relaxation rate $\Gamma$ as a function of $q^2$ at $time = 55$. (c) and (d) represent relaxation time $\tau$ and KWW as functions of time, correspondingly. Orange circles display results for CCS analysis, light blue squares - for ACS, and dashed blue line shows results from corresponding equilibrium systems. Results for other $q$ and $time$ values are similar. }
 \label{fgr:cuts_lin_decrease}
\end{figure}

\referencelist{}
\end{document}